\documentclass[twocolumn,prl,superscriptaddress,10pt,a4paper]{revtex4}
\usepackage{graphicx,amsmath,amsfonts,algorithm}
\floatname{algorithm}{Protocol}
\usepackage{graphics}
\usepackage{amssymb}
\usepackage{amsmath, amsthm}
\usepackage{complexity}
\usepackage{color}
\usepackage{bbm}

\newtheorem{prop}{Proposition}\def\PRO{\begin{prop}}\def\ORP{\end{prop}}
\newtheorem{coro}{Corollary}\def\COR{\begin{coro}}\def\ROC{\end{coro}}
\newtheorem{theo}{Theorem}\def\TH{\begin{theo}}\def\HT{\end{theo}}
\def\TH{\begin{theo}}\def\HT{\end{theo}}
\newtheorem{defi}[prop]{Definition}\def\DE{\begin{defi}}\def\ED{\end{defi}}
\newtheorem{lemme}[prop]{Lemma}\def\LE{\begin{lemme}}\def\EL{\end{lemme}}
\newcommand{\AR}[2][c]{$$\begin{array}[#1]{lllllllllllllll}#2\end{array}$$}

\def\EQ#1{\begin{eqnarray}#1\end{eqnarray}}

\def\bra#1{\langle#1{|}}

\def\op#1{\hat{#1}}
\def\ket#1{| #1 \rangle}

\def\bra#1{\langle #1 |}

\def\op#1#2{|#1\rangle\!	\langle#2|}

\def\dm#1{\op{#1}{#1}}
\def\ora#1{\overrightarrow{#1}}
\newcommand{\djj}{d\kern-0.4em\char"16\kern-0.1em}

\renewcommand{\section}[1]{{\em #1}.---}

\newcommand{\sect}[1]{ \vspace{0.1cm} {\large\bf #1}}

\begin{document}
\title{Blind Quantum Computing with Weak Coherent Pulses}

\author{Vedran Dunjko}
\affiliation{SUPA,
School of Engineering and Physical Sciences,
Heriot-Watt University, Edinburgh EH14 4AS, U.K.
}
\affiliation{Division of Molecular Biology, Ru\djj er Bo\v{s}kovi\'{c} Institute, Bijeni\v{c}ka cesta 54, P.P. 180, 10002 Zagreb, Croatia}

\author{Elham Kashefi}
\affiliation{
School of Informatics,
The University of Edinburgh, Edinburgh EH8 9AB, U.K.
}

\author{Anthony Leverrier}
\affiliation{ICFO - The Institute of Photonic Sciences Av. Carl Friedrich Gauss, num. 3, E-08860 Castelldefels (Barcelona), Spain}
\affiliation{Institute for Theoretical Physics, ETH Zurich, 8093 Zurich, Switzerland}
\begin{abstract}

The Universal Blind Quantum Computation (UBQC) protocol \cite{FOCS} allows a client to perform quantum computation on a remote server. In an ideal setting, perfect privacy is guaranteed if the client is capable of producing specific, randomly chosen single qubit states.  While from a theoretical point of view, this may constitute the lowest possible quantum requirement, from a pragmatic point of view, generation of such states to be sent along long distances can never be achieved perfectly.

We introduce the concept of $\epsilon$-blindness for UBQC, in analogy to the concept of $\epsilon$-security developed for other cryptographic protocols, allowing us to characterize the robustness and security properties of the protocol under possible imperfections. We also present a \emph{remote blind single qubit preparation} protocol with weak coherent pulses for the client to prepare, in a delegated fashion, quantum states arbitrarily close to perfect random single qubit states. This allows us to efficiently achieve $\epsilon$-blind UBQC for any $\epsilon>0$, even if the channel between the client and the server is arbitrarily lossy. 
\end{abstract}
\maketitle

While modern advances in quantum information are making strides towards scalable quantum computers, the dream of small and privately owned quantum computers remains very distant. Realistically, large quantum servers may in the near future take a role similar to that occupied by massive superclusters today. They will be remotely accessed by a large number of clients, using their home-based simple devices, to solve tasks which seem difficult for classical computers, while enjoying full privacy guaranteed by an efficient cryptographic scheme.

A similar question has been also addressed in the context of classical client-server scenarios. Considering unconditional security, Abadi, Feigenbaum and Killian studied the notion of ``computing with encrypted data'', and showed that no \NP-hard function can be computed blindly if unconditional security is required, unless the polynomial hierarchy collapses at the third level~\cite{AFK}. A related question which offers privacy with computational assumptions, known as fully homomorphic encryption, remained open for 30 years \cite{Gentry09}.

Various protocols have been devised with the goal of realizing such delegated, yet private and secure, quantum computing \cite{Childs,AS06,FOCS,Dorit,MDK11}. These protocols vary upon their requirements for the client and the achievable level of security. Among them, the universal blind quantum computing (UBQC) proposed by Broadbent, Fitzsimons and Kashefi \cite{FOCS} stands as the optimal one, with the lowest requirements on the client: in particular, no quantum memory is needed. The security offered by the ideal UBQC protocol is unconditional: the server cannot learn anything about the client computation, input or output. This flavour of security is called \emph{blindness}.
The feasibility of UBQC using different physical resources has been addressed \cite{MDK11,TomoFujii} and the potential of UBQC already prompted experimental demonstrations on a small scale \cite{ExpUBQC}.

The only `non-classical' requirement for the client in the ideal UBQC is that she can prepare single qubits in the state $\ket{+_{\theta}}\!=\!\frac{1}{\sqrt{2}}\! \left(\ket{0} + e^{i \theta} \ket{1} \right)$ with $\theta \in \{0, \pi/4,\ldots, 7\pi/4\}$. The blindness of the UBQC protocol has only been established in the ideal case where the client prepares perfect qubits. In any physical implementation, however, the preparation will inevitably be imperfect and this has to be taken into account before making any statement about security.
For instance, the qubits could be encoded in the polarization of a single photon generated by a realistic single photon source. Then completely suppressing the probability of inadvertently sending two or more identically polarized photons instead of one is very difficult, yet such an event would invalidate the perfect privacy of the client. While the future may bring scalable and fault-tolerant quantum computation required for the server, \textit{perfect} quantum devices required to guarantee the \textit{perfect} security for the client are unlikely to ever be achieved in practice.

The main contribution of this paper is towards this direction: we investigate the security of UBQC with realistic imperfections for the client. For this purpose, we introduce the framework of approximate blindness ($\epsilon$-\emph{blindness}) where the (small) parameter $\epsilon$ quantifies the maximal probability of successfully distinguishing between the actual protocol and the ideal one introduced in \cite{FOCS}. A similar approach to defining approximate security has been a milestone in the context of cryptography, and recently extended to the case of quantum key distribution (QKD) \cite{Ben-Or,scarani,MR09} and other quantum cryptographic primitives in the, so-called, bounded storage and noisy storage models \cite{BSM,koenig-2009,wehner07d}.

We will show that the level of security is indeed higher when the states prepared by the client are closer to the ideal qubits. We then introduce a protocol allowing the client to prepare the qubits in a delegated fashion at the server's location. The client needs to encode the quantum information into the polarization of weak coherent pulses which are sent to the server through an arbitrarily lossy quantum channel. Therefore, the burden of preparing very good qubits is put on the server, who needs in particular to be able to perform non-demolition quantum measurements \cite{QND}. We will show that these realistic requirements (for the client) are compatible with $\epsilon$-blind UBQC, where $\epsilon$ can be made arbitrarily small. This approach shares striking similarities with the history of QKD where the initial protocols required true single photons but later became compatible with the much more practical weak coherent pulses.

We begin with a brief recap of the UBQC protocol.

\section{Universal Blind Quantum Computation}
The UBQC protocol is set in the framework of measurement-based quantum computation (MBQC) \cite{RB01,RBB03,Jozsa05}. In MBQC the underlying resource is a  multipartite entangled quantum state and the computation is executed by performing measurements on its subsystems. In particular, this resource state can be a generic \emph{brickwork} state, a close relative of the cluster state (see the Appendix section B for details). By applying single qubit measurements parametrized by a \emph{measurement angle} $\phi$ from the discrete set $\{0, \pi/4, \ldots, 7\pi/4\}$,  which collapse the measured qubit state to one of the two eigenstates $\{\ket{\pm_\phi} = \frac{1}{\sqrt{2}}\! \left(\ket{0} \pm e^{i \phi} \ket{1} \right) \}$, with corresponding eigenvalues $\pm 1$, one can achieve universal quantum computation \cite{FOCS}.
Here, the computation itself is encoded in the measurement angles alone and the underlying resource is generic.

The classical and quantum part of MBQC can be conceptually separated.
One can imagine a classical controller which generates the measurement angles and a quantum unit which prepares the resource state, performs the measurements (as dictated by the controller unit), and returns the measurement outcomes to the controller unit. The outcomes are crucial for the \emph{adaptive structure} of MBQC: since they are probabilistic, the subsequent measurement angles must depend on them to ensure deterministic computation \cite{Flow06,gflow07}.

The central idea behind UBQC is to use this separation and allocate the classical controller unit 
to the client and the quantum unit to the server. To ensure privacy, however, the computation needs to be encoded: this is achieved in UBQC by effectively encoding the resource state.

The standard procedure for MBQC, to prepare the resource state, is to start with a set of qubits in a fixed state, say $\ket{+}:=\ket{+_0}$, and to apply an entangling operation of the Controlled Phase gate ($\textup{ctrl-}Z$) to some of them.

In UBQC, by contrast, the client will provide the initial phase rotated qubits of the form $\ket{+_\theta}$ to the server, without informing him of the values of $\theta \in \{0, \pi/4, \ldots, 7\pi/4\}$. Applying the entangling gates then prepares an encoded resource state. Now, if one was to measure a qubit in the usual MBQC protocol with some measurement angle $\phi$, this would be equivalent to measuring the pre-rotated qubit  (in the state $\ket{+_\theta}$ with the angle $\delta^\prime = \phi + \theta \mod\ 2\pi,$ as the phase rotation and $\textup{ctrl-}Z$ gate commute.
In this case, the measurement angle alone says nothing about the computation run, but a malicious server may still try learn something about $\theta$ when given $\delta^\prime$, hence also about $\phi$ (\emph{i.e.} about the computation).

To solve this security loophole, UBQC exploits the probabilistic nature of MBQC. The client sends a modified measurement angle  $\delta = \phi + \theta + r \pi \mod\ 2\pi,$ where $r\in \{0,1\}$ is chosen randomly by the client and hidden from the server.
The value of $r$ can be interpreted as a flip of the measurement outcome, which can be easily compensated by the client.

Now the quantum information (pre-rotated qubits) and classical information (measurement angles) accessible to the server is no longer correlated to the client's desired computational angles (denoted $\phi$), and this constitutes the crux of the proof of blindness of UBQC \cite{FOCS}.

One can summarize the UBQC protocol as follows:
Initially, in the \textit{preparation phase}, the client sends $S$ (the \emph{size} of the computation) randomly pre-rotated qubits in the states $\{\ket{+_{\theta_i}}\}_{i=1}^S$, to the server, keeping the angles $\theta_i$ secret. 
The server then builds up the brickwork state using the received qubits and the ctrl-$Z$ interaction. 
Proceeding sequentially on each qubit, if the desired measurement angle for qubit $i$ was $\phi_i$ (defined for the non-prerotated resource state, and including the necessary adaptations to the angle based on prior measurement outcomes $s_{k<i}$), the client will ask the server to measure the qubit with respect to the angle  $\delta_i = \phi_i + \theta_i + r_i \pi \mod\ 2\pi,$ where the binary parameter $r_i$ is chosen randomly.
The server reports each measurement outcome $s_i$ which the client flips if $r_i=1$.

In the case of an honest server, this procedure yields the correct outcome of the computation. Moreover, regardless of the malicious activity of the server the client's privacy is unconditional - the protocol is blind (see the Appendix section B for details). 

This blindness, however, only holds if the client can prepare the needed qubits perfectly. In a practical implementation, imperfection is inevitable and perfect blindness cannot be achieved. For this reason, a notion of \emph{approximate} blindness is required.

\section{Approximate blindness}
A difficulty to characterize the UBQC protocol is that it is adaptive. However, as far as blindness is concerned, the reported outcomes $s_i$ of the server do not matter; they only affect the correctness of the protocol 
\footnote{ After the pre-rotated qubits have been sent from the client to the server (and $\theta_i$ rotations fixed), the possible transcripts of the communication between the client and the server depend on two sequences of parameters: the measurement results $s_i$ sent by the server, and the (random) parameters $r_i$ chosen by the client. The classical information sent by the client depends on $r_i + s_i \mod\, 2$  (see step 3.5 of \textbf{Protocol 1} in the the Appendix, section  B).
 Since the parameters $r_i$ are random and unknown to the server, so are the values $r_i + s_i \mod\, 2$. Hence the responses from the client are independent from the choice of the servers reporting strategy. So we may fix the reporting strategy (and the measurement outcomes) without loss of generality.}. Hence one can assume $s_i=0$ (\emph{i.e.} the server measurement always projects into the $+1$ eigenvalue, similar to a post-selection scenario), and since the random parameters $r_i$ can be chosen in advance, the need for the adaptive structure can be ignored. Therefore, blindness can be studied through the following joint state of the client and server: 
\EQ{\label{idealState}
\pi_{AB} ^{\mathrm{ideal}}\!  = \dfrac{1}{2^{4S}}\! \sum\limits_{\ora{\phi}, \ora{r}}  \bigotimes_{i \in \left[S\right]}\! \underbrace{\dm{\phi_i} \otimes \dm{r_i}}_{\textrm{Client (A)}} \otimes 
 \underbrace{\dm{+_{\theta_i}} \otimes \dm{\delta_i}   }_{\textrm{Server (B)}},\nonumber}
  which contains all the relevant information pertaining to the security of a run of a UBQC protocol as seen by the server. The client has also access to the angles $\theta_i$.  However, these angles do not constitute the secret the client wishes to hide - for blindness only the $r_i$ parameters and the (adapted) computation angles are relevant. Thus in the presented joint state we explicitly place the client's secret in the client's register, and all information accessible to the server in the server's register.
    In this classical-quantum (cq) state, $S$ denotes the overall \emph{size} of the computation. The client's register contains the user's secret classical information - the computational angles $\phi_i$  characterizing  the desired computation, and the $r_i$ parameters chosen randomly and unknown to the server.
 The server's register contains quantum information - the qubits in states $\ket{+_{\theta_i}}$ which are sent by the client, as well as the measurement angles $\delta_i$. Note that $\phi_i, r_i$ and $\delta_i$ are all represented by classical, orthogonal states. 
 
If the information shared by the client and the server can be described by the state $\pi_{AB} ^{\mathrm{ideal}}$, then a malicious server cannot learn anything about the computation of the client \cite{FOCS}. Since the security holds for any action of the server, any UBQC protocol described by a state of the form  
 \EQ{\label{elem}
 (\mathbbmss{1}_{A} \otimes \mathcal{E}) \pi_{AB} ^{\mathrm{ideal}}
 }
 for any completely positive trace preserving map $\mathcal{E}$ (representing any possible deviation from the protocol by the server) is equally blind. We refer to such states as \emph{unconditionally blind states} and define the family $\mathcal{F}$ of such states as follows: 
  \EQ{ 
  \mathcal{F} = \left\{ (\mathbbmss{1}_{A} \otimes \mathcal{E}) \pi_{AB} ^{\mathrm{ideal}} \vert \mathcal{E}\textup{\ is\ a\ CPTP\ map} \right\} .}
 
 In order to analyse the impact of imperfections caused by a realistic implementation, we consider the settings where the client sends general states $\rho^{\theta_i}$ instead of the perfect states $ \ket{+_{\theta_i}}$. In this case, the joint state representing all the information exchanged in the protocol is given by:
\EQ{\label{leaky}
\pi_{AB} ^{\{\rho^{\theta_i} \}}  = \dfrac{1}{2^{4S}} \sum\limits_{\ora{\phi}, \ora{r}}  \bigotimes_{i \in \left[S\right]} \underbrace{\dm{\phi_i} \otimes \dm{r_i}}_{\textrm{Client}} \otimes 
 \underbrace{\rho^{\theta_i} \otimes \dm{\delta_i}   }_{\textrm{Server}}.
 }

We can now introduce the notion of $\epsilon$-blindness:
\DE
A UBQC protocol with imperfect client preparation described by the shared joint state $\pi_{AB} ^{\{\rho^{\theta_i} \}} $ is $\epsilon$-blind if the trace distance between the family of unconditionally blind states and the state $\pi_{AB} ^{\{\rho^{\theta_i} \}} $ is less than $\epsilon$:
\EQ{
\min_{ \pi_{AB}^{\mathcal{E}} \in \mathcal{F}} \frac{1}{2}  \| \pi_{AB}^{\{\rho^{\theta_i }\}} - \pi_{AB}^{\mathcal{E}} \|  \leq \epsilon \, \label{gencrit} 
}
\ED

Such a notion of security is particularly desirable as it is composable \cite{2005-3607,PhysRevLett.98.140502, Ben-or05theuniversal, MR09}. One can also extend it to a more general setting considering prior knowledge about the computation (see the Appendix, section C).

If the states $\rho^{\theta_i}$ generated by the client are uncorrelated (which holds for instance if the process determining the parameters $\theta_i$ is random and memoryless), the distance between the perfectly blind state and the approximate state $\pi_{AB} ^{\{\rho^{\theta_i} \}} $ can be bounded in terms of the distance between the individual states $\rho^{\theta_i}$ and the corresponding perfect qubit states $\ket{+_{\theta_i}}$.
In particular, defining
\EQ{
  \epsilon_{\mathrm{prep}} = \max_{\theta_i} \frac{1}{2} \| \rho^{\theta_i} - \mathcal{E}(\dm{+_{\theta_i}}) \| \label{constr}
} 
 for some CPTP map $\mathcal{E}$ independent of all $\theta_i$, then one can show (see the Appendix, section C) that
\EQ{
\min_{ \pi_{AB}^{\mathcal{E}} \in \mathcal{F}} \frac{1}{2}  \| \pi_{AB}^{\{\rho^{\theta_i}\}} - \pi_{AB}^{\mathcal{E}} \|  \leq  S \epsilon_{\mathrm{prep}} \label{criterion1}. 
}
This means that the ability to prepare good approximations of the states $\ket{+_{\theta_i}}$ translates into the ability to perform approximately-blind universal quantum computing.

This, however, is not completely satisfying from the client's perspective. Indeed, the client can only achieve a given value of $\epsilon_{\mathrm{prep}}$ in practice, meaning that for a fixed security parameter $\epsilon$, she cannot perform a computation with more that $\epsilon/\epsilon_{\mathrm{prep}}$ steps.
In order to allow for computation of arbitrary size, it is necessary to prepare arbitrary good qubits and the solution is to delegate this task to the server, who is assumed to be much more powerful than the client. 

We proceed by presenting such a \textit{Remote Blind qubit State Preparation} (RBSP) protocol where the client only needs to prepare weak coherent pulses with a given polarization. The requirements for the client are therefore minimal. In particular, they are the same as in most practical implementations of discrete-variable QKD. The difficulty here is transferred to the server who has to perform a quantum non-demolition measurement to obtain the desired qubit. As we will show, using the RBSP protocol $S$ times, the client can reach a joint state $\pi_{AB} ^{\{\rho^{\theta_i} \}} $ which is  $\varepsilon$-close to the family $\mathcal{F}$ of perfectly blind states.

\section{UBQC with Remote Blind qubit State Preparation using weak coherent pulses}
The RBSP protocol is designed to serve as a substitute for the process of sending one individual perfect random qubit which allows for imperfect devices and channel.

Ideally, its outcome will satisfy the following properties: \textbf{(A)} the state in the server's possession is $\mathcal{E}(\dm{+_\theta})$ for a CPTP map $\mathcal{E}$, independent of $\theta$ known to the client alone --  guaranteeing perfect blindness, see Eq. (\ref{constr});  \textbf{(B)} the protocol is never aborted in the honest server scenario -- guaranteeing robustness of the encompassing UBQC; \textbf{(C)} in the honest server scenario, the map $\mathcal{E}$ is the identity --  guaranteeing the correctness of the UBQC protocol.

When imperfections are taken into account, the UBQC using RBSP in the preparation phase approaches the properties of blindness and robustness asymptotically. Hence, we are interested in the following properties: $\epsilon$-blindness, as described above, and $\epsilon$-robustness which guarantees that the honest abort probability is less than $\epsilon$. Despite the imperfect preparation stage, we also show that the correctness of the protocol holds in the honest scenario, whenever the client does not abort.

To run the RBSP protocol, the client sends a sequence of $N$ weak coherent pulses (small amplitude, phase-randomized coherent states) with random polarization $\sigma$ in the set $\{0, \pi/4,\ldots, 7\pi/4\}$ to the server. If the transmittance of the channel from the client to the server is supposed to be at least $T$, then the mean photon number of the source is set to $\mu=T$. This value of $\mu=T$ is optimal for our security analysis, however other values are in principle admissible as well. 
The introduced phase randomization simplifies the security analysis and causes the state emitted from the source to be:
\EQ{
\rho^{\sigma} = \sum_ {k=0}^\infty p_k \ket{k}\!\bra{k}_{\sigma} \nonumber
}
where $\ket{k}_{\sigma} := |+_\sigma\rangle^{\otimes k}$ corresponds to $k$ photons, occurring with probability $p_k=e^{-T}T^k/k!$, with polarization $\sigma$.
Each pulse is then a probabilistic mixture of Fock states. The Poissonian distribution obtained here is not crucial for the
RBSP protocol. For instance, it would work equally well (with
re-adjustment of parameters) with any source realizing a mixture of
polarization encoded photon number states, such as polarized
thermal states, provided that the probability of getting a single
photon is not too small.

The server then performs non-demolition photon number measurements on the pulses he receives, declaring the number outcomes to the client. This  additional requirement on the quantum server, while a challenging task has already been experimentally implemented \cite{haroche}.
At this point, the client checks the number of reported vacuum states - if this number is greater than $ N(e^{-T^2} + T^2/6)$, she aborts the protocol. A higher value would be indicative to either a lossier than believed channel, or more importantly, that the server lied in an attempt to cheat.

If the protocol was not aborted, the server performs the interlaced 1D cluster computation subroutine (I1DC), using the photons obtained by the number measurement of the received coherent pulses. 
In this subroutine, the server couples the first and the second qubit (\emph{i.e.} photons) with the interaction $\textup{ctrl-}Z.(H\otimes \mathbbmss{1})$, and the first qubit of the pair is then measured in the Pauli $X$ basis and the measurement outcome is sent to the client. The remaining qubit is then coupled to the third qubit in the input set and measured in the same basis. This process is repeated until only one qubit remains unmeasured, in some state $\ket{+_{\theta}}$.

Using her knowledge about the polarizations of each of the pulses initially sent, and the reported binary string of outcomes, the client can compute the angle $\theta$ (see the Appendix section {\bf D} for details). The pair $\theta$ (held by the client) and $\ket{+_\theta}$ (held by the server), is the required outcome of the RBSP protocol. 

The intuition behind this protocol is the following. The I1DC subroutine is such that if the server is totally ignorant about the polarization of at least one photon in the 1D cluster, then he is also totally ignorant about the final angle $\theta$. In order to exploit this property, the client should make sure that the server will at least once measure a single photon and put it in the cluster. The cheating strategy for the server consists in claiming he received 0 photon when he received 1 and claiming he received 1 when he in fact measured several (in which case he can learn something about their polarization). In order to avoid this attack, the client simply verifies that the reported statistics of the server are compatible with the assumed transmittance of the channel. 
Note that the server cannot learn anything useful, even if he deviates from the prescribed I1DC subroutine if one of the weak coherent pulses generated one photon, and was declared as such.
We now give more quantitative statements which are proven in the Appendix (together with detailed descriptions of both RBSP and I1DC).

For the described RBSP protocol, property \textbf{(A)} holds except with probability $p_{\mathrm{fail}}$ and properties \textbf{(B)} and \textbf{(C)} hold except with probability $p_{\mathrm{abort}}$. 
These probabilities $p_{\mathrm{abort}}$ and $p_{\mathrm{fail}}$ can be bounded as functions of the transmittance $T$ and the parameter $N$ as follows:
\begin{equation}
p_{\mathrm{fail}}, \, p_{\mathrm{abort}} \leq \exp\left(-\frac{N T^4}{18} \right) \,.\label{bound}
\end{equation} 

Using the bound on $p_{\mathrm{fail}}$, the trace distance between the perfectly blind qubit state and the state $\rho_{\theta}$ generated by RBSP can be bounded as
\begin{equation}
\dfrac{1}{2}\left\| \rho^{\theta}   -  \mathcal{E}(\dm{+_\theta}) \right\| \leq p_{\mathrm{fail}} \nonumber
\end{equation}
for a fixed CPTP map $\mathcal{E}$ independent of $\theta$.
From this, by the criterion given in expression (\ref{criterion1}),  the bound given in Eq. (\ref{bound}) and  the union bound, we have that a protocol using the RBSP generated states is $\epsilon$-blind with $\epsilon \leq S \exp\left(-N T^4/18 \right),$ where $T$ is a lower bound on the channel transmittance, and $N$ the number of states used in each instance of RBSP.
These results are proven in details in the Appendix and collected in the following main theorem:
\TH\label{main}
A UBQC protocol of computation size $S$, where the client's preparation phase is replaced with $S$ calls to the coherent state Remote Blind qubit State preparation protocol, with a lossy channel connecting the client and the server of transmittance no less than $T$, is correct, $\epsilon$-robust and $\epsilon$-blind for a chosen $\epsilon>0$ if the parameter $N$ of each instance of the Remote Blind qubit State preparation protocol called is chosen as follows:
$$
N \geq  \dfrac{18\ln(S/\epsilon)}{T^4}.
$$
\HT

We acknowledge that the RBSP protocol is not immune to noise in the channel or to significant preparation errors on the side of the client. A method of performing RBSP in a fault tolerant way, by adapting techniques used to ensure the fault tolerance of UBQC itself \cite{TomoFujii, FOCS,TopCode},  is under the investigation of the authors. However, noise can only jeopardize the correctness of our protocol, but never the guaranteed security levels.

\section{Conclusions and outlook}
In this work we have addressed the security of UBQC under the presence of imperfections through the concept of $\epsilon$-blindness. Following this we have given a Remote qubit State Preparation pre-protocol which allows a client, with access to weak coherent pulses only, to enjoy UBQC with arbitrary levels of security.

The transition from the idealized setting of UBQC using single photon qubits to the present protocol using weak coherent pulses brings UBQC significantly closer to real-life applications for, e.g. an unconditionally secure quantum network. The parallel with the evolution of QKD is also very interesting.
Note that in QKD, weak coherent pulses were not very attractive for long distance communication before the invention of protocols with decoy states \cite{decoy}. Indeed, these decoy states made the optimal intensity $\mu$ of the attenuated laser be roughly a constant, in contrast with the optimal $\mu \approx T$ without decoy states. In the case of UBQC, it might also be the case that decoy states could improve the optimal value of $\mu$ and therefore significantly decrease the required number weak coherent pulses used in an instance of RBSP for a given computation.

\section{Acknowledgements}
We thank Erika Andersson for insightful discussions. We would also like to acknowledge the hospitality of the Telecom ParisTech Quantum Group where this work was initiated during the visits by all the authors.  VD is supported by EPSRC (grant EP/G009821/1), EK is supported by EPSRC (grant EP/E059600/1) and AL received financial support from the EU ERC Starting grant PERCENT. This work was done while AL was at ICFO.

\pagebreak

\begin{widetext}
{\LARGE {\bf Appendix} }\\
\vspace{0.1cm}

\noindent\sect{A Outline}\\

In section B 
, we give a detailed description of the UBQC protocol, the definition of blindness and the statement of the main theorem guaranteeing blindness of UBQC from \cite{FOCS}. Section 
{\bf C}
gives the definitions of $\epsilon$-blindness in presence of prior knowledge about the computation and the derivation of the expression (6) from the main paper. This expression links the trace distance between ideal prepared qubits and imperfectly prepared states with the blindness parameter $\epsilon$ of the encompassing UBQC which uses these imperfect states. In Section D
, we give a detailed description of the I1DC subroutine, with the proof of its correctness, and in Section E
, we present the details of the entire Remote Blind State Preparation (RBSP) protocol. We prove the security characteristics of RBSP claimed in the main paper. Finally, in the last section, we derive the characterization of the states generated be RBSP in terms of the criterion given by the expression (5) from the main paper, which is the crux of our main security theorem for UBQC with weak coherent pulses.\\

\noindent\sect{B Universal Blind Quantum Computation}\\

\label{UBQC}

For the detailed overview of the UBQC protocol, we will assume the familiarity with the measurement-based quantum computing, for more details see \cite{RB01,Mcal07}. Suppose the client has in mind a unitary operator $U$ that is implemented with a measurement pattern on a brickwork state $\mathcal{G}_{n\times m}$ (Figure~\ref{fig:cluster}) with measurements given as multiples of~$\pi/4$ in the $(X,Y)$ plane with overall computation size $S = n \times m.$
This pattern could have been designed either directly in MBQC or from a circuit
construction.
Each qubit $\ket{\psi_{x,y}} \in  \mathcal{G}_{n \times m}$ is
indexed by a \emph{column} $x \in \{1, \ldots ,n\}$ and a
\emph{row}~$y \in \{1, \ldots ,m\}$. Thus each qubit is assigned a measurement
angle~$\phi_{x,y}$, a set of $X$-dependencies $D_{x,y} \subseteq [x-1]\times[m]$ and
a set of $Z$-dependencies $D'_{x,y} \subseteq [x-1]\times [m]$\,.
Here, we assume that the dependency sets~$X_{x,y}$ and~$Z_{x,y}$ are obtained via the
flow construction~\cite{Flow06}.

During the execution of
the pattern, the actual measurement angle $\phi'_{x,y}$ is computed from $\phi_{x,y}$ and the previous measurement
outcomes in the following way: let $s^X_{x,y} = \oplus_{i\in
D_{x,y}}{s_i}$ be the parity of all measurement outcomes for qubits
in $X_{x,y}$ and similarly, $s^Z_{x,y} = \oplus_{i\in D'_{x,y}}{s_i}$ be
the parity of all measurement outcomes for qubits in~$Z_{x,y}$.

Then
$ \phi'_{x,y} = (-1)^{s^X_{x,y}} \phi_{x,y} + s^Z_{x,y} \pi$\,. 

{\bf Protocol \ref{prot:UBQC}} implements a blind quantum computation for $U$. 

We assume that the client's input and output of the computation are built into $U$. In other words, the client wishes to compute the results of some fixed single qubit measurements in the $(X,Y)$ plane of the state $U (\ket{+}\ldots\ket{+})$. Note however, that the protocol could be easily extended to deal with arbitrary classical or quantum input and output \cite{FOCS}.

\renewcommand{\labelenumi}{\textbf{\arabic{enumi}.}}
\renewcommand{\labelenumii}{\arabic{enumi}.\arabic{enumii}}
\renewcommand{\labelenumiii}{\arabic{enumi}.\arabic{enumii}.\arabic{enumiii}}

\begin{algorithm}
\caption{Universal Blind Quantum Computation}
 \label{prot:UBQC}

\begin{enumerate}

\item  \label{step:client-prep} \textbf{Client's preparation} \\
For each column $x = 1, \ldots , n$, \\
\hspace*{\parindent}\hspace*{\parindent} for each row $y = 1,
\ldots , m$,
\begin{enumerate}
\item \label{step:one-one} the client prepares the state $\ket{\psi_{x,y}}
 \in  \{ \ket{+_{\theta_{x,y}}}:= \frac{1}{\sqrt{2}}(\ket{0} + e^{i\theta_{x,y}}\ket{1}) \mid \theta_{x,y} = 0,
\pi/4, \ldots, 7\pi/4 \}$, where the defining angle $\theta_{x,y}$ is chosen uniformly at random, and sends the qubits to the server.
\end{enumerate}

\item  \label{step:server-prep} \textbf{Server's preparation} %
\begin{enumerate}
\item The server creates an entangled state from all received qubits, according to their indices,
by applying \textsc{ctrl}-$Z$ operators between the qubits in order to
create a brickwork state $\mathcal{G}_{n \times m}$.
\end{enumerate}

\item  \label{step:computation-interaction} \textbf{Interaction and measurement}
\\
For each column $x = 1, \ldots , n$ \\
\hspace*{\parindent}\hspace*{\parindent} For each row $y = 1,
\ldots , m$
\begin{enumerate}
\item \label{step:three-one} The client computes $\phi'_{x,y}$
where $s^X_{0,y}=s^Z_{0,y}=0$.
\item The client chooses a binary digit $r_{x,y} \in \{0,1\}$ uniformly at random, and computes $\delta_{x,y} =
\phi'_{x,y}  + \theta_{x,y} + \pi r_{x,y}$.
\item The client  transmits $\delta_{x,y}$ to the server, who performs a measurement in the basis $\{ \ket{+_{\delta_{x,y}}},
\ket{-_{\delta_{x,y}}} \}$.
\item The server transmits the result $s_{x,y} \in \{0,1\}$ to the client.
\item \label{one-timepad}If $r_{x,y} = 1$, the client flips $s_{x,y}$;
otherwise she does nothing.
\end{enumerate}
\end{enumerate}
\end{algorithm}

 \begin{figure}[h] \centering
   \includegraphics[scale=0.8]{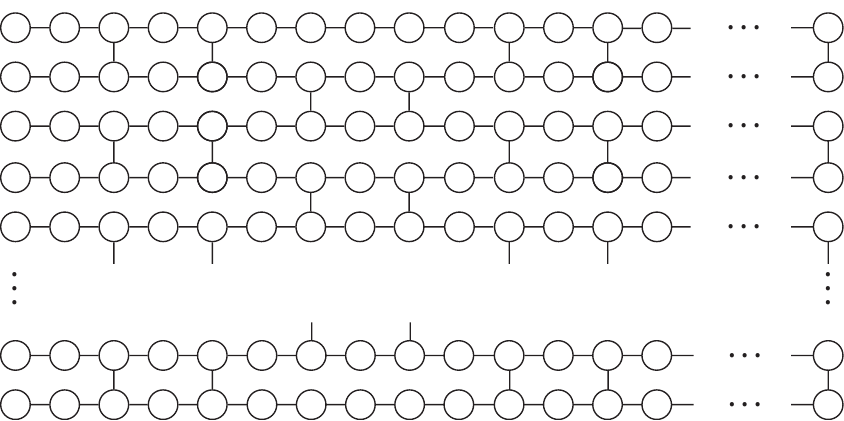}

 \caption{{The \emph{brickwork state}, $\mathcal{G}_{n \times m}$, a universal resource state for measurement-based quantum computing requiring only single qubit measurement in the $(X,Y)$ plane \cite{FOCS}.
    Qubits $\ket{\psi_{x,y}}$ $(x=1, \ldots, n, y=1,\ldots,m)$ are arranged according to layer $x$ and row $y$,
     corresponding
 to the vertices in the above graph, and are originally in
 the~$\ket{+} = \frac{1}{\sqrt{2}}\left(\ket{0}
 +\ket{1}\right)$ state.
    Controlled-$Z$ gates are then performed
     between qubits which are joined by an edge.
    The rule determining which qubits are joined by an edge is as follows: 1) Neighbouring qubits of the same row are joined; 2) For each column $j = 3 \mod\, 8$ and each odd row $i$, the qubits at positions $(i, j)$ and $(i + 1, j)$ and also on positions $(i, j + 2)$ and $(i + 1, j + 2)$ are joined; 3) For each column $j = 7 \mod\, 8$ and each even row $i$, the qubits at positions $(i, j)$ and $(i + 1, j)$ and also on positions $(i, j + 2)$ and $(i + 1, j + 2)$ are joined. }
    \label{fig:cluster}}
 \end{figure}

In the main text and in this Appendix, for simplicity, we have used single indexing for all the involved parameters: the measurements angles $\phi_i$, the random variables $r_i$ and $\theta_i$, the messages to the server $\delta_i$ (characterising the measurement angles to be performed) and the server's messages to the client $s_i$ (measurement outcomes). In particular, one has $\delta_i = (-1)^{s^X_{i}} \phi_{i} + s^Z_{i} \pi + \theta_i + r_i\pi$. It was shown in \cite{FOCS} that this protocol is correct, \emph{i.e.}, if both the client and the server follow the steps of the protocol then the classical outcome are the results of some fixed single qubit measurements of the state $U(\ket{+ \ldots +}$). These classical outcome corresponds to the signals $s_{i}$ generated by the measurements of the final layer of the brickwork state then bit flipped if the corresponding parameter $r_{i}$ was 1 and left as they are if the corresponding $r_{i}$ was zero.
Moreover, the server will not learn anything about the client's computation, \emph{i.e.}, the protocol is \emph{blind} with respect to the following definition:

\begin{defi} \label{defn:blind} We say a protocol \textsf{P} on input $X$ is \emph{blind while leaking at most L(X)}, where $L(X)$ is any function of the input if:
\begin{enumerate}
\item \label{defn:blind-1} The distribution of the classical information obtained by the server in \textsf{P} depends only on $L(X)$.
\item Given the distribution of classical information described in~\ref{defn:blind-1} and $L(X)$, the state of the quantum system obtained by the server in \textsf{P} is fixed.
\end{enumerate}
\end{defi}

\begin{theo}\cite{FOCS}\label{thm:privacy}
{\bf Protocol~\ref{prot:UBQC}} is blind while leaking at most the dimensions of the brickwork state, i.e. an upper bound on the input size and the depth of the computation.
\end{theo}

Equivalently, blindness implies that in {\bf Protocol \ref{prot:UBQC}}, from the server's point of view, the distribution of the computational angles, $\phi_i$, is uniform and the final classical output is one-time padded \footnote{If $b=(b_1, \ldots, b_n)$ and  $k=(k_1, \ldots, k_n)$ are two bit strings of equal length, then the bit string defined as $\tilde{b} = (b_1 \oplus k_1, \ldots, b_n \oplus k_n)$ is the one-time padded bit string $b$ (using the key $k$). The significance of the one-time pad (also known as the Vernam cipher) is that an adversary intercepting the one-time padded message $\tilde{b}$ cannot learn anything about the message $b$ unless he has the access to the key $k$, and is the crux of essentially all information-theoretically secure cryptosystems. In the quantum setting consider a general $n$ qubit state $\rho$ and define the state $\rho^\prime =\left(\bigotimes_{i=1}^{n} Z_i^{z_i}X_i^{x_i} \right) \rho \left(\bigotimes_{i=1}^{n}  X_i^{x_i} Z_i^{z_i} \right) $. The subscripts on the Pauli operators $X$ and $Z$ denote which qubit they act upon, and the superscripts $x_i$ and $z_i$ are random secret keys, designating whether the corresponding Pauli operator is applied or not. We say that the state $\rho^\prime$ is the (quantum) one-time padded state, using the keys ${(z_i,x_i)}_{i=1}^n.$
If an adversary has the state $\rho^\prime$, without the knowledge about the one-time pad keys, he cannot gain any information about the state $\rho$. This is easy to see as in this case the state seen by the adversary is a totally mized state, $\frac{1}{4^n} \sum\limits_{i=1}^n \sum\limits_{x_i,y_i =0}^{1} \left(\bigotimes_{i=1}^{n} Z_i^{z_i} X_i^{x_i} \right) \rho \left(\bigotimes_{i=1}^{n}  X_i^{x_i} Z_i^{z_i} \right)$.}, see \cite{FOCS} for the proof. Moreover, in the setting where the server's prior knowledge exists, the server does not acquire any new knowledge about the angles $\phi_i$ through the UBQC protocol.\\

\noindent\sect{C Approximate blindness with prior knowledge}\\

\label{Approx-prior}
In UBQC, the unitary transformation $U$ that the client wishes to perform is encoded with a vector of angles $\ora{\phi} = (\phi_1, \ldots, \phi_S)$. In general, it is reasonable to consider the scenario where the server may have prior information about the transformation $U$ the client wishes to run.
Since, in the framework of UBQC, the computation is encoded in the angles $\phi_i$  the server's prior knowledge can be modelled by assigning a non-uniform \emph{a priori} probability distribution $p(\ora{\phi})$. If the delegated computation protocol is blind, then the posterior distribution for the server should be equal to the prior one. 
This was established for the original UBQC protocol \cite{FOCS}.

While the UBQC protocol is inherently adaptive, one can show that the $s_i$ values do not contribute towards the blindness purposes and only effect the correctness of the protocol. 
Recall that after the initial qubits have been sent from the client to the server (i.e. having fixed the `pre-rotation' angles $\theta_i$), the possible transcripts of the communication between the client and the server depend on two sequences of binary parameters: the declared measurement results $s_i$ sent by the server, and the (random) choices of the parameters $r_i$ chosen by the client. The classical information sent by the client is easily seen to depend on $r_i + s_i \mod\, 2$  (see step 3.5 of \textbf{UBQC protocol}).
Since the parameters $r_i$ are chosen at random and unknown to the server, so are the values $r_i + s_i \mod\, 2$. Hence every reporting strategy chosen by the server will result in responses from the client which are independent from the choice of the servers reporting strategy. So we may fix the reporting strategy (and the measurement outcomes) without the loss of generality.

In what follows we define $\epsilon$-blindness for a UBQC with imperfect client preparation and prior knowledge. For the more special case where there is no prior knowledge, addressed in the main text, one need just set the probability distribution $p(\ora{\phi})$ to be uniform.
Blindness can be studied through the following joint ideal blind state, with \emph{a priori} probability distribution $p(\ora{\phi})$ and relative to the CPTP map $\mathcal{E}$,  defined as
\EQ{
\pi_{AB} ^{\mathcal{E}, \, p(\ora{\phi})}  = \mathbbmss{1}_{\mathrm{Client}} \otimes \mathcal{E}_{\mathrm{Server}} \; \left( \dfrac{1}{2^{S}} \sum\limits_{\ora{\phi}, \ora{r}} p(\ora{\phi}) \bigotimes_{i \in \left[S\right]} \underbrace{\dm{\phi_i} \otimes \dm{r_i}}_{\textrm{Client}} \otimes 
 \underbrace{\left(\dm{+_{\theta_i}} \otimes \dm{\delta_i} \right)  }_{\textrm{Server}} \right )\,. \nonumber
 }
 
 which corresponds to the  entirety of information, pertinent to the blindness of a UBQC protocol run, the client and the server share in the post-selected setting where all the signals $s_i$ are set to 0. This is, as we noted, done without the loss of generality.
Additional information the client and the server have access to would include the classical descriptions of the angles $\theta$.  However, these angles do not constitute the secret the client wishes to hide. The server may learn something about these angles, but what blindness requires is that this knowledge reveals nothing about the computational angles, or the hidden $r$ parameters. The presented ideal state is suitable for addressing this question, by explicitly placing the client's secret in the client's register, and all information accessible to the server in the server's register.
 
 Here, the map $\mathcal{E}$ is introduced because the server might deviate from the original protocol by applying such a map, as we have done in the setting with no prior knowledge in our paper. However, because he cannot get any information about the client's secret by performing such an operation, the final state remains as blind as in the absence of this map.
Correspondingly, we define the family $\mathcal{F}^{p(\ora{\phi})}$ of ideal blind states with respect to prior knowledge $p(\ora{\phi})$ as follows:

\EQ{
\mathcal{F}^{p(\ora{\phi})} = \left\{  \pi_{AB} ^{\mathcal{E}, \, p(\ora{\phi})}\ \vert \ \mathcal{E}\ \text{is a CPTP map} \right\} \,. \nonumber
}

Any protocol characterized by a state in $\mathcal{F}^{p(\ora{\phi})}$ is as secure as the ideal UBQC protocol with prior knowledge $p(\ora{\phi})$. Again, like in the setting with no prior knowledge, the correctness of the protocol can be guaranteed only if the map $\mathcal{E}$ is the identity. In general, the state describing a given protocol does not correspond to an ideal state, but is given by the following expression:

\EQ{\label{imperfectprior}
\pi_{AB}^{\rho^{\{\theta_i\}_i},~p(\ora{\phi})} =   \dfrac{1}{2^{S}} \sum\limits_{\ora{\phi}, \ora{r}} p(\ora{\phi})  \bigotimes_{i \in \left[S\right]} \underbrace{\dm{\phi_i} \otimes \dm{r_i}}_{\textrm{Client}} \otimes 
 \underbrace{\rho^{\theta_i} \otimes \dm{\delta_i}   }_{\textrm{Server}}.
 }

\DE
A UBQC protocol with imperfect client preparation and prior knowledge $p(\ora{\phi})$, in which the client sends states of the form $\rho^{\theta_i}$ instead of the perfect states $\ket{+_{\theta_i}}$ in the preparation phase is called an \emph{$\epsilon$-blind UBQC protocol with imperfect states and prior knowledge $p(\ora{\phi})$}, if the trace distance between the overall joint state given with the expression (\ref{imperfectprior}) and  the family $\mathcal{F}^{p(\ora{\phi})}$ of unconditionally blind joint states with prior knowledge is less than $\epsilon$:

\EQ{
\min_{\pi_{AB}^{\mathcal{E},p(\ora{\phi})} \in \mathcal{F}^{p(\ora{\phi})}} \dfrac{1}{2}  \| \pi_{AB}^{\rho^{\{\theta_i\}_i},p(\ora{\phi})} - \pi_{AB}^{\mathcal{E},p(\ora{\phi})} \| \leq \epsilon \,. \nonumber
}

\ED
The above criterion is equivalent to
\EQ{
\min_{\mathcal{E}} \dfrac{1}{2} \| \pi_{AB}^{\rho^{\{\theta_i\}_i},~p(\ora{\phi})} - \pi_{AB}^{\mathcal{E},p(\ora{\phi})} \| \leq \epsilon \nonumber
}
where the map $\mathcal{E}$ ranges over all CPTP maps.

This notion of approximate security makes a crucial use of the trace distance between the state obtained while running the actual protocol and an ideal state. This approach has been already used in the context of quantum cryptography and particularly for quantum key distribution, where the $\epsilon$-security of a protocol is defined analogously. The importance of the trace distance comes from the fact that it is closely linked to the maximal probability of distinguishing the actual protocol from the ideal one. Therefore, if this probability is arbitrary small, then the actual protocol is arbitrary secure.

Next we compute how an approximate preparation of the states $\ket{+_{\theta_i}}$ affects the blindness of the overall protocol. We wish to bound the distance 
\EQ{
\min_{\mathcal{E}}  \| \pi_{AB}^{\rho^{\{\theta_i\}_i},~p(\ora{\phi})} - \pi_{AB}^{\mathcal{E},\, p(\ora{\phi})} \| \nonumber
}
where the minimization is over all possible CPTP maps $\mathcal{E }$  acting on the system in the possession of the server.
First, we can restrict ourselves to the maps $\mathcal{E}$ which act individually and identically on the subsystems containing the qubits $\dm{+_{\theta_i}}$
\EQ{
\min_{\mathcal{E}} \! \left\| \pi_{AB}^{\rho^{\{\theta_i\}_i},~p(\ora{\phi})}\! -\! \pi_{AB}^{\mathcal{E},\,p(\ora{\phi})} \right\| \!\leq\! 
\min_{\mathcal{E}, \mathrm{i.i.d.}}\!  \left\| \pi_{AB}^{\rho^{\theta_i},~p(\ora{\phi})}\! -\! \pi_{AB}^{\mathcal{E},\,p(\ora{\phi})
}\right\|\,. \nonumber
}
This clearly holds as the minimization of the right-hand side of the expression above is just the minimization restricted to the subset of the minimization space of the left-hand side of the expression.
One has

\begin{eqnarray}
 \left\| \pi_{AB}^{\rho^{\{\theta_i\}},p(\ora{\phi})} - \pi_{AB}^{\mathcal{E},p(\ora{\phi})} \right\| 
 &= & \dfrac{1}{2^{S}}
\left\|   \sum\limits_{\ora{\phi}, \ora{r}} p(\ora{\phi}) \left( \bigotimes_{i \in \left[S\right]} \dm{\phi_i}  \dm{r_i}  
  \rho^{\theta_i} \dm{\delta_i} - \bigotimes_{i \in \left[S\right]} \dm{\phi_i}  \dm{r_i} 
 \mathcal{E}(\dm{+_{\theta_i}})  \dm{\delta_i}  \right) \right\|  \nonumber\\
 &\leq&  \dfrac{1}{2^{S}} \sum\limits_{\ora{\phi}, \ora{r}} p(\ora{\phi}) \left\| \bigotimes_{i \in \left[S\right]} \dm{\phi_i}  \dm{r_i}  
   \rho^{\theta_i}  \dm{\delta_i}        - \bigotimes_{i \in \left[S\right]} \dm{\phi_i}  \dm{r_i}  
\mathcal{E}(\dm{+_{\theta_i}})\dm{\delta_i}\right\|  \nonumber \\
&\leq& \dfrac{1}{2^{S}} \sum\limits_{\ora{\phi}, \ora{r}} p(\ora{\phi}) \left\| \bigotimes_{i \in \left[S\right]}  \rho^{\theta_i}   -     \bigotimes_{i \in \left[S\right]} \mathcal{E}(\dm{+_{\theta_i}})\right\| \nonumber\\
&\leq &   \dfrac{1}{2^{S}}  \sum\limits_{\ora{\phi}, \ora{r}} p(\ora{\phi}) \sum_{i \in \left[S\right]} \left\|   \rho^{\theta_i}   -  \mathcal{E}(\dm{+_{\theta_i}})\right\| .\label{20} 
\end{eqnarray}

Although  the variables $\theta_i$ are drawn uniformly at random initially, their distribution, given $\delta_i$ and some prior knowledge about the angles $\phi_i$ is not uniform. While the expression $(\ref{20})$ could be refined further, this general derivation becomes rather cumbersome, and is omitted here. Now we can characterize the quality of the qubit preparation by the parameter $\epsilon_\mathrm{prep}$ defined as:
\AR{
 \epsilon_\mathrm{prep} = \displaystyle{\max_{\theta}} \dfrac{1}{2} \left\|   \rho^{\theta}   -  \mathcal{E}(\dm{+_{\theta}})\right\|.  \nonumber
}
We then obtain:
\AR{
\left\| \pi_{AB}^{\rho^{\{\theta_i\}_i},p(\ora{\phi})} - \pi_{AB}^{\mathcal{E},p(\ora{\phi})} \right\|&\leq &   \dfrac{1}{2^{S}}  \sum\limits_{\ora{\phi}, \ora{r}} p(\ora{\phi}) \sum_{i \in \left[S\right]} \left\|   \rho^{\theta_i}   -  \mathcal{E}(\dm{+_{\theta_i}})\right\| \\
 &\leq &  \dfrac{1}{2^{S+1}}  \sum\limits_{\ora{\phi}, \ora{r}} p(\ora{\phi}) S  \epsilon_\mathrm{prep} = 2 S  \epsilon_\mathrm{prep} \label{21}.
}
Hence, in such a scenario, the obtained UBQC protocol is $\epsilon$-blind if:
\begin{equation}
\label{condition2}
S  \epsilon_{\mathrm{prep}} \leq \epsilon \, .
\end{equation}
This proves the bound given in Eq. (6) in the main text. \\

\noindent \sect{D Interlaced 1D cluster computation}\\

\label{I1DC:sect}
The Interlaced 1D cluster computation protocol, used as a subroutine in RBSP, is described in \textbf{Protocol \ref{I1DC}}.

\begin{algorithm}
\caption{Interlaced 1-D Cluster computation (I1DC)}
\label{I1DC}
\begin{itemize}
\item \textbf{Input:} A sequence of $k$ states $(\ket{+_{\sigma_l}})_{l=1}^k$ for $\sigma_l \in_R \{\frac{j \pi }{4}\}_{j=0}^7$.
\item \textbf{Output:} A binary string of measurement outcomes $s=(s_1, \ldots, s_k )$ and the state $\ket{+_\theta},$ where \EQ{
\theta = \sum\limits_{l=1}^k (-1)^{t_l} \sigma_l \, , \label{theta}}
where the binary components $t_i$ are given as follows:
 \EQ{
 t_i = 
 \left\{
     \begin{array}{ll}
         \sum_{j=i}^{k-1} s_i \mod 2,  & \text{for} \quad i < k\\
 		  0 &\text{for} \quad  i = k\\
     \end{array}\label{ts}
 \right.
 }
\item \textbf{Computation steps:}

\begin{enumerate}
\item For $i=1$ to $(k-1)$
\begin{enumerate}
\item Apply the unitary $\textup{ctrl-Z}(H\otimes \mathbbmss{1})$ to qubits $i$ and $i+1$.
\item Measure qubit $i$ in the Pauli-$X$ basis, obtaining the outcome $s_i$.
\end{enumerate}
\item{Return the string $s = (s_1, \ldots, s_k)$ and the remaining non-measured qubit in the state $\ket{+_\theta}$.}

\end{enumerate}
\end{itemize}
\end{algorithm}

In the RBSP protocol, an honest server will follow the presented I1DC subroutine exactly. He will have sent the measurement outcomes $s= (s_1, \ldots, s_k)$ to the client who will then compute the angle $\theta$ using the formulas $\ref{theta}$ and $\ref{ts}.$ For the correctness of the encompassing RBSP protocol, we need to show that the angle the client computes is the same as the angle parametrizing the state kept by the server. This constitutes the correctness of the I1DC protocol, and we give it with the following lemma:

\LE\label{I1DCLemma}
\textbf{Protocol \ref{I1DC}} is correct.
\EL
\begin{proof}
We will prove by induction that the state of the output qubit in the interlaced 1-D computation protocol performed on the input of $k$ qubits in the states $\{\ket{+_{\sigma_l}}\}_{l=1}^k$, given the sequence of measurement outcomes $(s_1, \ldots, s_{k-1})$ is the state $\ket{+_\theta},$ where \EQ{
\theta = \sum\limits_{l=1}^k (-1)^{t_l} \sigma_l}
where the binary parameters $ (t_1, \ldots, t_k)$ are computed as follows: 
\EQ{
t_i = 
\left\{
    \begin{array}{ll}
        \sum_{j=i}^{k-1} s_i \mod 2,  & \text{for} \quad i < k,\\
		  0 &\text{for} \quad  i = k.\\
    \end{array}
\right. 
}

For the basis of the induction we verify that the claim holds for the first non-trivial case, $k=2$.
Consider the state 
\EQ{
\textup{ctrl - Z} (H \otimes \mathbbmss{1}) \ket{+_{\sigma_1}} \otimes \ket{+_{\sigma_2}} \nonumber
}
It is easy to check that the state of the second subsystem, after the measurement of the Pauli-X observable on the first subsystem of the state
is the state $\ket{+_{\sigma_2 + (-1)^{s_1} \sigma_1}},$
where $s_1 = 0$ corresponds to the measurement outcome associated to the post-measurement state $\ket{+}$ and $s_1 = 1$ to the outcome associated to the post-measurement state $\ket{-}$. Then, according to Equation (\ref{ts}), $t_1 = s_1$ and $t_2 = 0$, and then Equation $(\ref{theta})$ gives
\EQ{
\theta = (-1)^{t_1} \sigma_1 + (-1)^{t_2} \sigma_2 =  (-1)^{s_1} \sigma_1 + \sigma_2, \nonumber
}
which is the angle corresponding to the resulting state for the case $k=2.$

Assume then the step of the induction, \emph{i.e.} that the claim holds for the input size $k=n$, and let us then show that it then also holds for $k=n+1$. Consider the case where the computational steps of the I1DC protocol have been run to the $n^{th}$ step, i.e. to finish off the protocol, the output qubit of the first $n$ steps of the computation needs to be entangled to the $(n+1)^{st}$ qubit using the prescribed interaction and measured in the Pauli $X$ eigenbasis. Let $(s_1, \ldots, s_{n-1})$ be the measurement outcomes of the first $n-1$ measurements. Then by the step of the induction the state of the output qubit of the first $n$ steps is $\ket{+_{\theta^\prime}}$ where
\EQ{
\theta^\prime = \sum\limits_{l=1}^n (-1)^{t_l ^\prime} \sigma_l \nonumber
}
and
\EQ{
t_i^\prime = 
\left\{
    \begin{array}{ll}
        \sum_{j=i}^{n-1} s_i \mod 2,  & \text{for} \quad i < n,\\
		  0 &\text{for} \quad  i = n.\\
    \end{array}
\right. \nonumber
}
If the entangling interaction is then applied on this resulting qubit $\ket{+_{\theta^\prime}}$ and the remaining qubit $\ket{+_{\sigma_{n+1}}}$, and the first qubit is then measured in the Pauli $X$ eigenbasis, by the basis of the induction, the resulting state is $\ket{+_\theta}$ where:
\EQ{
\theta = (-1)^{s_n} \theta^\prime +  \sigma_{n+1}. \nonumber
} 

This in turn can be expanded as 
\begin{eqnarray*}
\theta &=& (-1)^{s_n} \theta^\prime +  \sigma_{n+1} \\
&=& (-1)^{s_n}\sum\limits_{l=1}^n (-1)^{t_l ^\prime} \sigma_l +  \sigma_{n+1}\\
&=& (-1)^{s_n}\sum\limits_{l=1}^n (-1)^{ \left(\sum_{j=i}^{n-1} s_i \mod 2 \right)} \sigma_l +  \sigma_{n+1}\\
 & =& \sum\limits_{l=1}^n (-1)^{ \left(\sum_{j=i}^{n} s_i \mod 2 \right)} \sigma_l +  \sigma_{n+1} \\
 &=&  \sum\limits_{l=1}^{n+1} (-1)^{t_l} \sigma_l 
\end{eqnarray*}
for 
\EQ{
t_i = 
\left\{
    \begin{array}{ll}
        \sum_{j=i}^{n-1} s_i \mod 2,  & \text{for} \quad i < n+1,\\
		  0 &\text{for} \quad  i = n+1.\\
    \end{array}
\right.\nonumber
}
Hence, the I1DC protocol is correct.
\end{proof}

\noindent\sect{E Remote Blind qubit State Preparation}\\

\label{RBSP:sect}

\begin{algorithm}
\caption{Remote Blind qubit State Preparation with weak coherent pulses with parameters $(N, T)$}
\label{RBSP2}
\begin{enumerate}
\item \textbf{Client's preparation}
\begin{enumerate}
\item The client generates $N$ weak coherent pulses with mean photon number $\mu = T$ and a randomized phase and a polarisation $\sigma_l$ (for $l = 1, \ldots, N$). These states are described by
\begin{equation}
\rho^{\sigma_l} =  e^{-\mu} \sum_{k=0}^\infty \frac{\mu^{k}}{k!} |k\rangle\!\langle k|_{\sigma_l}.
\end{equation}
The polarisation angles $\sigma_l$ are chosen uniformly at random in $\left\{k \pi/4 : 0 \leq k \leq 7 \right\}$. The client stores the sequence $(\sigma_1, \ldots, \sigma_N).$

\item The client sends the states $\{\rho^{\sigma_l}\}_l$ to the server.
\end{enumerate}
\item \textbf{Server's preparation}
\begin{enumerate} 
\item For each state he receives, the server performs a non-demolition measurement of the photon number, obtaining a sequence of $N$ classical values and $N$ post-measurement states. If the measured photon number was greater than zero, the server keeps one photon, discarding the rest. 

\item The server reports the string  $(n_1, \ldots, n_N)$ to the client.
\end{enumerate}
\item \textbf{Client-server interaction}
\begin{enumerate}

\item The client verifies that the reported number of vacuum states is not too large with respect to the tolerated value of the transmittance $T$ of the quantum channel between the client and the server. More precisely, if this number is larger than $N (e^{-T^2} + T^2/6)$, then the client aborts the protocol. 

 Otherwise the protocol continues.

\item The server discards the systems for which he measured zero photon. Each subsystem with $n_l>0$  photons measured, parametrized by the polarisation $\sigma$, is interpreted as a system of $n_l$ qubits in the state $\ket{+_{\sigma_l}}.$ Only one qubit copy per received state is kept, and the total remaining number of qubits is $M$.

\item Using the qubits from the step above and respecting the sending order, the server performs the I1DC computation (see {\bf Protocol \ref{I1DC}}), obtaining the sequence $t=(t_1, \ldots, t_M)$ and keeping the resulting state $\ket{+_\theta}$.

\item The server reports the string $t$ to the client.

\item Using her knowledge about the angles $\sigma_l$ of the qubits used in the I1DC procedure by the server, and the received outcome string $t$, the client computes $\theta$ with formula (\ref{theta}).
\end{enumerate}
\end{enumerate}
\end{algorithm}

The Remote Blind qubit State Preparation protocol (RBSP) is described by \textbf{Protocol \ref{RBSP2}}. In order to prover the security characteristics of RBSP, we prove the following properties which together with the properties of the original UBQC protocol in \cite{FOCS} prove the claims in the main text:
\begin{itemize}
\item[\textbf{(A)}] Upon the completion of the RBSP protocol the state in the server's possession is $\mathcal{E}(\dm{+_\theta})$ for some CPTP map $\mathcal{E}$ (independent of $\theta$) and the client alone knows the angle $\theta$, except with probability $p_{\mathrm{fail}}$;  
\item[\textbf{(B)}] The protocol is never aborted in the honest server scenario, except with probability $p_{\mathrm{abort}}$;
\item[\textbf{(C)}] In the honest server scenario,  the map $\mathcal{E}$ is the identity if the client did not abort and the protocol is correct.
\end{itemize}
The correctness of the protocol in property $\textbf{(C)}$ above means that upon the successful completion of the RBSP protocol, the server has the state $\dm{+_\theta}$ where $\theta$ is the angle the client has computed.

We claim that he probabilities $p_\mathrm{fail}$ and $p_\mathrm{abort}$ are bounded above in terms of the protocol parameter $N$, and relative to the transmittance lower bound $T$ as follows:
\begin{equation}
p_{\mathrm{fail}}, \, p_{\mathrm{abort}} \leq \exp\left(-\frac{N T^4}{18} \right) \nonumber.
\end{equation}

\begin{proof}
We begin by proving \textbf{Claim (C)}, which is a consequence of the correctness of the interlaced 1-D computation protocol, Lemma \ref{I1DCLemma}.
For \textbf{Claim (C)} to hold, first it needs to be shown that, if the protocol was not aborted, and the server is honest, then the server's system is in the state $\ket{+_{\theta}}$ for some $\theta$ known only to the client.  In the case where the server is honest, prior to the call to the interlaced 1D cluster computation subroutine, the server's system is in the state $$\bigotimes_{l=1}^k \ket{+_{\sigma_k}}$$ where the angles $\sigma_k$ are known to the client.
Then the server will perform the interlaced 1D cluster computation using this system as the input, reporting the bit string $(t_1, \ldots, t_k)$, which is related to the measurement outcomes, as explained in \textbf{Protocol \ref{I1DC}}. The client will then compute the angle $\theta$ using the formula $(\ref{theta})$. Hence, by Lemma \ref{I1DCLemma} this angle is precisely the angle defining the state  $\ket{+_{\theta}}$ in the server's subsystem.
 What remains to be seen is that the angle of this resulting state is chosen uniformly at random. Recall that the angle $\theta$ is given with $\theta = \sum\limits_{l=1}^k (-1)^{t_l} \sigma_l = \sum\limits_{l=1}^{k-1} (-1)^{t_l} \sigma_l + \sigma_{k}$, and since the angles $\sigma_{k}$ are polarisation angles and they are assumed to be chosen uniformly at random, the angle $\theta$ is also distributed uniformly at random. 
This proves that the client alone knows the value of $\theta.$

To prove \textbf{Claim (B)}, we will need to bound the abort probability when the server is honest. 
Finally, for  \textbf{Claim (A)}, we will need to show that if the protocol is not aborted then the state in the server's possession is $\mathcal{E}(\dm{+_\theta})$ for some CPTP map $\mathcal{E}$, where $\theta$ is the angle the client will compute based on the servers feedback, except with probability $p_\mathrm{fail}$. 

We address these two required properties throughout the rest of this section. In Lemma \ref{lemma}, which we present later, we show that if during the run-time of the protocol the server measures a single photon in one of the states (coherent pulses) sent by the client and declares it as such, then if the client does not abort the protocol, the resulting state with the server is $\mathcal{E}(\dm{+_\theta})$ for a CPTP map $\mathcal{E}$, where $\theta$ is the angle the client will compute based on the servers feedback. Hence, the probability of this not happening, is the failure probability, $p_\mathrm{fail}$. 

Here, we note $p_k = e^{-\mu}\frac{\mu^{k}}{k!}$ the probability of receiving $k$ photons if the channel is perfect (unit transmittance), and $p_k^T=e^{-T \mu} \frac{(T \mu)^k}{k!}$ the probability of receiving $k$ photons if the quantum channel between the client and the server is a lossy channel of transmittance $T$. 
In fact, since the events with 2 or more photons are not distinguished by our protocol, we note $p_{\geq 2}$ (resp. $p_{\geq 2}^T$) the probability of obtaining 2 or more photons for a perfect channel (resp. a channel with transmittance $T$). 

In what follows, we derive the bounds for both $p_{\mathrm{fail}}$ (blindness) and $p_{\mathrm{abort}}$ (robustness). For each state that the server receives, he is supposed to perform a non demolition measurement of the photon number and to announce this number to the client.
Here, we are only interested in three types of events: 
\begin{itemize}
\item ``event 0'' when the server measures 0 photon. This event has probability $p_0^T$ in the case of an honest server since the transmission channel is characterized by a transmittance $T$. 
\item ``event 1'', when the server measures exactly 1 photon. The whole point of the protocol is to make sure that at least once, this event occurs and the server has to announce that he received one photon. If this is the case, by Lemma \ref{lemma} we are guaranteed the server has the desired state $\mathcal{E}(\dm{+_\theta})$ for a CPTP map $\mathcal{E}$.
\item ``event 2'', when the server measures at least 2 photons. In the case of a malicious server, one has to suppose that the probability of such an event is $p_{\geq 2}$ (instead of $p_{\geq 2}^T$), meaning that we assume that the server has the ability to replace the imperfect quantum channel by a lossless one.
\end{itemize}

As we will explain below, without loss of generality we may assume that the server always performs the number measurement. Then, if the server is malicious, his only strategy consists in declaring he received 0 photon when he detected 0, declaring he received 0 photon when he detected 1, and declaring either 1 or more when he detected at least 2 photons. Any other strategy will either mean that the server will admit to having measured one photon in which case, by Lemma $\ref{lemma}$ the protocol will end in a satisfactory state. Alternatively, the server has to report that he measured 1 photon when he in fact measured none. In this case in the setting where the client did not abort, the angle the client computes will be uncorrelated to the state generated by the server.  However, this corresponds to a state of the form $\mathcal{E}(\dm{+_\theta})$ for a CPTP map $\mathcal{E}$ where the map $\mathcal{E}$ is the contraction to the state in the server's possession, independent of the client's calculated angle $\theta$.
We will prove this formally in Lemma \ref{lemma2} presented later.

Let us denote with $N$ the total number of states sent by the client, $M_0, M_1$ and $M_2$ the number of states for which the server \emph{measured} respectively 0, 1 or at least 2 photons. Also define $N_0, N_1$ and $N_2$ to be the respective numbers of states for which the server \emph{reported} having measured 0, 1 or at least 2 photons.
Note that the numbers $M_0, M_1$ and $M_2$ are well-defined since the server does not gain anything by not measuring the photon number for each state he receives. This is because the measurement operators commute with the state sent by the client, which are diagonal in the Fock basis. We can therefore assume that he performs this non-demolition measurement.

These various quantities are related through the normalization constraint
\begin{equation}
M_0 + M_1 + M_2 = N_0 + N_1 + N_2 = N. \nonumber
\end{equation}
For an honest server, one has $N_0 = M_0, N_1 = M_1, N_2 = M_2$. A malicious server will however choose a strategy such that $N_0 = M_0 + M_1$. Consider the probability that the protocol aborts when the server is honest.
Hoeffding's bound \cite{citeulike:3392582} immediately gives an upper bound for $p_{\mathrm{abort}}$, for any $\Delta > 0$ we have
\begin{eqnarray}
p_{\mathrm{abort}} &=& \mathrm{Pr}\left[\frac{M_0}{N} - p_0^T \geq \Delta \right]\\
&\leq & \exp (- 2 \Delta^2 N). \nonumber
\end{eqnarray}

The only way the protocol fails is that the malicious server applies the strategy described above, (that is, to pretend he did not receive anything unless he actually received at least two photons) while not being detected. Let us consider a tolerance $\Delta$ which will be optimized later. One has:
\begin{eqnarray}
p_{\mathrm{fail}} &=& \mathrm{Pr}\left[\frac{N_0}{N} - p_0^T \leq \Delta \right] \nonumber\\
&\leq & \mathrm{Pr}\left[\frac{M_0 + M_1}{N} - p_0^T \leq \Delta \right] \nonumber\\
&\leq & \mathrm{Pr}\left[1 - \frac{M_2}{N} - p_0^T \leq \Delta \right] \nonumber\\
&\leq & \mathrm{Pr}\left[\frac{M_2}{N} - p_2  \geq 1 - p_0^T -p_{\geq 2} -\Delta \right] \nonumber\\
&\leq & \exp (- 2 \tilde{\Delta}^2 N),
\end{eqnarray}
with
\begin{equation}
\tilde{\Delta} := 1 - p_0^T -p_{\geq 2} - \Delta.\nonumber
\end{equation}
In order to get a non-trivial bound for $p_{\mathrm{fail}}$, the parameter $\tilde{\Delta}$ should be positive and bounded away from 0.
One has
\begin{eqnarray}
\tilde{\Delta} + \Delta &=& 1 - e^{-T \mu} - (1 - (1+\mu)e^{-\mu}) \nonumber \\
&=& e^{-\mu} \left( 1+ \mu - e^{(1-T)\mu} \right) \, .\nonumber
\end{eqnarray}
If we fix $\mu=T$, we obtain
\begin{eqnarray}
\tilde{\Delta} + \Delta = e^{-T} \left( 1+ T - e^{T(1-T)} \right)
\geq  \frac{T^2}{3} \,.\nonumber
\end{eqnarray}
Hence, choosing $\Delta = \tilde{\Delta} \geq T^2/6$, one gets
\begin{equation}
p_{\mathrm{fail}}, \, p_{\mathrm{abort}} \leq \exp\left(-\frac{N T^4}{18} \right) \,.\nonumber
\end{equation} \end{proof}

While the probabilities $p_{\mathrm{fail}}$ and $p_{\mathrm{abort}}$ can in principle be made arbitrary small for any (positive) value of the transmittance, one notes that the required number of weak coherent pulses scales like $\log(1/\epsilon)/T^4$ for small $T$, making the scheme less efficient. In general, this subroutine will be used $S$ times during the complete UBQC protocol. The probabilities that blindness or robustness is jeopardized during this whole process can be bounded easily with the union bound and they are simply increased by a factor $S$. This means that the correct scaling for the parameter $N$ should be $(\log (S/\epsilon)) / T^4$.

In the above, we considered one specific implementation of the remote blind qubit state preparation protocol using weak coherent pulses. This choice was made because weak coherent pulses are arguably the simplest quantum states to prepare in a laboratory. However, the protocol could be easily generalised to any source of light that emits a mixture of Fock states. In particular, the protocol would work equally well with a thermal source of light. The only characteristics which are required are that the probability of emitting exactly one photon is strictly positive, and that the client is able to calibrate her source well enough. In other words, reasonable bounds on the probability of emitting a given number of photons should be available.

Next, we present the lemmas we need to complete the proof above.

\LE \label{lemma}
If the server measured a weak coherent pulse sent by the client to contain one photon, declared it as such honestly to the client, and the client did not abort in the presented remote blind qubit state preparation protocol, then the state in the possession of the server after the termination of the protocol is $\mathcal{E}(\dm{+_\theta})$ for a CPTP map $\mathcal{E}$ where $\theta$ is the angle computed by the client.
\EL

\begin{proof}

We begin the proof by describing the system of the client and the server after the server has reported the binary string $\{t_i\}_i$ to the client, which he has to do to prevent the client from aborting. The client has the following:
\begin{itemize}
\item A sequence of angles $ \{ \sigma_k \}_{k=1}^{M}$ which the client has encoded in the polarization of the initially sent coherent pulses, corresponding to those pulses for which the server has announced a non-zero declared photon number. In this sequence the angles come in multiples, with individual indexes, the multiplicity corresponding to the announced number of photons declared. The total number of photons declared is then $M.$
\item A sequence of binary digits $\{t_k\}_k$ reported by the server, where the last digit $t_{M}$ is zero.
\end{itemize}

By assumption, the server measures at least one pulse for which he gets one photon and declares one photon. Without loss of generality, let us assume that this is the case for the final pulse characterized by its polarization angle $\sigma_M$.
The client will then calculate the value $\theta = \sum_{i=1}^{M-1} \sigma_i + \sigma_M$.

On the server's side, prior to declaring the binary digit outcomes, the server's quantum state can in all generality be written as:
\AR{
\eta^{\sigma_1, \ldots, \sigma_{M-1}} \otimes \dm{+_{\sigma_M}} \, ,
}
where the state $\ket{+_{\sigma_M}}$ is the state of the single copy declared qubit.
The rest of the server's system depends on the number measurement outcomes, but can always be written in the generic form $\eta^{\sigma_1, \ldots, \sigma_{M-1}}$.
In the case the server was honest, his system will be exactly of the form:
\AR{
\left( \dm{+_{\sigma_1}}\otimes \ldots \otimes  \dm{+_{\sigma_{M-1}}} \right) \otimes \dm{+_{\sigma_M}} \, ,
}
where the states before the final state may come in multiplicities which match the declared number measurement outcomes.

Note that, whatever procedure the server may run on the system in his possession, in the spirit of the Stinespring dilation theorem,f can always be represented as a unitary transform $U$ on the input system, augmented by an ancillary system, followed by a measurement on the output of the overall unitary transform. The classical outcome will, in general, encode the binary digits $\{t_k\}_k$ the server has to report to the client, as in the case of no report the protocol is aborted. We emphasize that we are not assuming anything about the classical outcome - it may be a result which depends on the states the server recieved from the client, it may be chosen randomly by the server, or it may be selected. The state in the server's possession prior to measurement, $\rho^{\mathrm{prior}}$, can then be viewed as the result of a CPTP map (which depends on the state $\eta^{\sigma_1, \ldots, \sigma_{M-1}}$) applied on the state  $\dm{+_{\sigma_M}}$
\AR{
\rho^{\mathrm{prior}} = \mathcal{E}^{\eta^{\sigma_1, \ldots, \sigma_{M-1}}} (\dm{+_{\sigma_M}}).
}
This is only possible because the angle $\sigma_M$ does not depend on any other angles. For simplicity, we shall fix the angles $\sigma_1, \ldots, \sigma_{M-1}$ and simply write the state prior to measurement: 
\AR{
\rho^{\mathrm{prior}} = \mathcal{E} (\dm{+_{\sigma_M}}).
}
Since the angle $\sigma_M$ was chosen uniformly at random, known to the client, the state of the server's system is: 
\AR{
\pi_\mathrm{server} = 1/8 \sum\limits_{\sigma_M} \mathcal{E}(\dm{+_{\sigma_M}}).
}

Following this, the server will measure a part of his subsystem, obtaining the sequence of binary digits $\ora{t}=\{t_k\}_k$ which he reports to the client. The state of the system after measurement (taking into account all possible outcomes) can be written as:
\AR{
\pi_\mathrm{server} = 1/8 \sum\limits_{\sigma_M} \sum\limits_{\ora{t}} p_{\sigma_M}(\ora{t})   \mathcal{E}_{\ora{t}}(\dm{+_{\sigma_M}}),
}
where $p_{\sigma_M}(\ora{t})$ is the probability of outcome $\ora{t}$ given that the input state was $\sigma_M$, and $\mathcal{E}_{\ora{t}}$ are the quantum operations which depend on the outcome.
Note that:
\EQ{
  \sum\limits_{\ora{t}} p_{\sigma_M}(\ora{t})\mathcal{E}_{\ora{t}} = \mathcal{E} \label{reflem2}
}
for all angles $\sigma_M.$
As the two sums commute, we can write this state as:
\EQ{
\pi_\mathrm{server}\! =\! 1/8\!  \sum\limits_{\ora{t}}\! \sum\limits_{\sigma_M}\! p_{\sigma_M}(\ora{t})\mathcal{E}_{\ora{t}}(\dm{+_{\sigma_M}}). \label{lemSubs}
}
Recall that a fixed sequence $\ora{t}$ along with the fixed sequence of angles $\sigma_1, \ldots, \sigma_{M-1}$ defines the angle $\theta$:
\AR{
\theta = \sum_{k=1}^{M-1} (-1)^{t_k} \sigma_{k} + \sigma_M.
}
Note that the value of $\theta$ attains all possible angles when $\sigma_M$ goes through all possible angles, for $\ora{t}$  and  $\sigma_1, \ldots, \sigma_{M-1}$ fixed. Now, since the sum:
\AR{
 \sum\limits_{\sigma_M} p_{\sigma_M}(\ora{t})   \mathcal{E}_{\ora{t}}(\dm{+_{\sigma_M}}) \, ,
}
for a fixed sequence $\ora{t}$ goes through all the possible angles, this sum is, for every sequence $\ora{t}$, equal to:
\AR{
\pi_\mathrm{server}= 1/8\!  \sum\limits_{\ora{t}}\!\sum\limits_{\theta} p_{\theta}(\ora{t})   \mathcal{E}_{\ora{t}}(\dm{+_{\theta}}).
}
Due to property (\ref{reflem2}), this final state is of the form $\mathcal{E}(\dm{+_\theta})$. \end{proof}

\LE\label{lemma2}
If the server measured a weak coherent pulse sent by the client to contain zero photon, and declared it as containing one photon to the client, and the client did not abort in the presented remote blind qubit state preparation protocol, then the state shared by the client and the server after the termination of the protocol is of the form $\mathcal{E}(\dm{+_\theta})$ for some CPTP map $\mathcal{E}$.
\EL

\begin{proof}
We will extensively use the setup and the arguments of the proof of Lemma \ref{lemma}. Assume that it is the $l^\mathrm{th}$ declared photon that the server does not possess. 
Then for the sequence of binary digits $(t_1 \ldots, t_k)$ the server will have reported as the alleged classical outcome of the interlaced 1D cluster computation, the angle the client computes is given as: 
\EQ{
\theta = \sum_{i=1}^{k} (-1)^{t_k} \sigma_{i}\, , \nonumber
}
which can be written as:
\EQ{
\theta = \sum_{i\in \{1, \ldots, l-1, l+1, \ldots k\}} (-1)^{t_i} \sigma_{i} + (-1)^{t_l} \sigma_l \,.\nonumber
}

Let us fix all the $\sigma_i$ angles except $\sigma_l$ and all the reported binary digits $t_i$ except $t_l$. The general state with the server after the remote blind state preparation protocol can then be written as:
\EQ{
\pi_\mathrm{server} = \sum\limits_{t_l=0}^{1} p(t_l) \sum_{\sigma_l} \frac{1}{8}  \eta^{t_l} \nonumber
} 
where $\eta$ is the state in the hands of the server, which may depend on $t_l$, and $p(t_l)$ is the probability of the server reporting $t_l$ to be one or zero.
Note that neither the probability $p(t_l)$ nor the final state $\eta^{t_l}$ can depend on $\sigma_l$.
The angle $\theta$ in the expression above for any fixed $t_l$ goes across all possible values as $\sigma_l$ ranges across all possible values. Hence the sum may be written in terms of the angle $\theta$ rather than $\sigma_l$ as: 
 \EQ{
 \pi_\mathrm{server} = \sum\limits_{t_l=0}^{1} p(t_l) \sum_{\theta}\frac{1}{8}\eta^{t_l} \, .\nonumber
 } 
 Also, since $\eta^{t_l}$ does not depend on $\sigma_l$ it does not depend on $\theta$ so we can factor it out of the sum
  \EQ{
  \pi_\mathrm{server} = \sum\limits_{t_l=0}^{1} p(t_l) \left( \sum_{\theta} \frac{1}{8}\right) \eta^{t_l} =\sum\limits_{t_l=0}^{1} p(t_l)  \eta^{t_l} \, .\nonumber
  } 
Let $\eta$ be the quantum state $\sum\limits_{t_l=0}^{1} p(t_l)  \eta^{t_l}$, then we have
 \EQ{
    \pi_\mathrm{server} =    \eta \nonumber
    } 
which implies 
 \EQ{
    \pi_\mathrm{server} =    \mathcal{E}^{\eta} (\dm{+_\theta}) \nonumber
    } 
where $\mathcal{E}^{\eta}$ is a CPTP map which is the contraction to the fixed state $\eta$.
\end{proof}

\noindent\sect{F Characterization of the states generated by RBSP}
\\

As we discussed before, the only source of imperfection in the implementation of the UBQC protocol comes from the fact that the client cannot generate exactly the states $|+_\theta\rangle$, but only an approximate state $\rho^\theta$. Using the remote blind qubit state preparation protocol with parameters $(N,T)$, the state generated instead of $|+_\theta\rangle$ can be described, in a worst-case scenario from the client's point of view, as
\begin{equation}
\rho^\theta = (1-p_{\mathrm{fail}}) \mathcal{E}^{S}(|+_\theta\rangle\langle +_\theta | )\otimes |0\rangle\langle 0|  + p_{\mathrm{fail}} |\theta\rangle\langle \theta|  \otimes |1\rangle\langle 1|, \label{endstate}
\end{equation}
where $p_\mathrm{fail}\leq \exp\left(-\frac{N T^4}{18} \right)$ and $|\theta\rangle$ is a classical state giving full information about $\theta$ \footnote{The state space of the sub-register containing the state $\mathcal{E}^{S}(|+_\theta\rangle\langle +_\theta | )$ or the state $ |\theta\rangle\langle \theta|$ above is of dimensionality no less than 8 (to be able to store the classical information) and both of the states should be thought of as being encoded using some orthonormal basis of the register. For instance, if the register state space is spanned by $\{\ket{k}\}_{k=0}^8$ then the state $\ket{\theta}$ could be encoded as $\ket{\theta}: = \ket{k}$ for $\theta = k \pi/4$ and the qubit state $\ket{+_\theta}$ as  $\ket{+_\theta}: = 1/\sqrt{2} (\ket{0} + \exp(i k \pi/4 ) \ket{1})$ }.
This state corresponds to the worst case scenario because we assume that if the server manages to avoid ``event 1'' (measuring a single photon in a weak coherent pulse and declaring it as such), corresponding to the indicator register with the state $|1\rangle\langle 1|$, which happens with probability $p_\mathrm{fail}$ at most, then the server obtains complete information about the angle $\theta$ chosen by the client. As we have shown, in any other cases the server has the state $\mathcal{E}^{S}(|+_\theta\rangle\langle +_\theta | )$ for some CPTP map $\mathcal{E}^S$, relative to the angle $\theta$ computed by the client. In modelling this, we considered an additional quantum register with values $\ket{0}$ or $\ket{1}$, available to the server, which tells whether the remote blind qubit state preparation succeeded or not, as it can be seen in Equation (\ref{endstate}). This is a classical-quantum state, and the classical register containing either  $\ket{0}$ or $\ket{1}$ represents the information whether the server managed or failed to cheat and the server clearly has access to this information.

We have that
\begin{equation}
\dfrac{1}{2}\left\| \rho^{\theta}   -  \mathcal{E}(\dm{+_\theta}) \right\| \leq p_{\mathrm{fail}} \nonumber
\end{equation}
for every $\theta$ and the CPTP map $\mathcal{E}(\rho):= \mathcal{E}^S(\rho) \otimes |0\rangle\langle 0| $. This map simply applies the CPTP map $\mathcal{E}^S$ to the input state and appends the state $\ket{0}.$
Hence, in order to characterize blindness, one just needs to consider the condition described in Equation (\ref{condition2}), that is $S \epsilon_{\mathrm{states}} \leq \epsilon$ and to note that
\begin{equation}
p_{\mathrm{fail}} \leq S  \exp\left(-\frac{N T^4}{18} \right).\nonumber
\end{equation}
The latter claim holds as we have taken into account that the remote blind qubit state preparation protocol is used $S$ times during the complete UBQC protocol and the probabilities that blindness or robustness is jeopardized during this whole process can be bounded easily with the union bound as they are simply increased by a factor $S$. Hence, choosing $N =  18\ln(S/\epsilon)/T^4$ allows the client to obtain $\epsilon$-blindness for the overall UBQC protocol for arbitrary small values of $\epsilon$. The above explanation, given the proven bounds on the probabilities $p_\mathrm{fail}$ and $p_\mathrm{abort}$,  proves the main theorem of our paper.
\end{widetext}

\end{document}